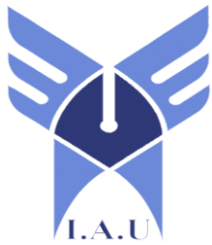
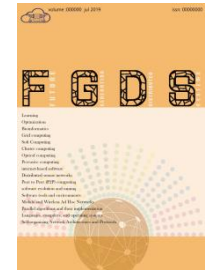

# Feature Selection-based Intrusion Detection System Using Genetic Whale Optimization Algorithm and Sample-based Classification

Amir Mojtahedi[1], Farid Sorouri[2], Alireza Najafi Souha[2], Aidin Molazadeh[3], Saeedeh Shafaei Mehr[1]

1- Department of Computer Engineering, Moghadas Ardabili Institute of Higher Education, Ardabil, Iran
Email: Amir7mojtahedi@gmail.com(Corresponding author)
Email: Shafaeemehrs@gmail.com
2- Department of Computer Engineering, Ardabil Branch, Islamic Azad University, Ardabil, Iran
Email: fa.sorouri@gmail.com
Email: a.najafi@lauardabil.ac.ir
3-National Industrial Group, Tehran, Iran
Email: Dr.molazadeh1@gmail.com

**ABSTRACT:**
Preventing and detecting intrusions and attacks on wireless networks has become an important and serious challenge. On the other hand, due to the limited resources of wireless nodes, the use of monitoring nodes for permanent monitoring in wireless sensor networks in order to prevent and detect intrusion and attacks in this type of network is practically non-existent. Therefore, the solution to overcome this problem today is the discussion of remote-control systems and has become one of the topics of interest in various fields. Remote monitoring of node performance and behavior in wireless sensor networks, in addition to detecting malicious nodes within the network, can also predict malicious node behavior in future. In present research, a network intrusion detection system using feature selection based on a combination of Whale optimization algorithm (WOA) and genetic algorithm (GA) and sample-based classification is proposed. In this research, the standard data set KDDCUP1999 has been used in which the characteristics related to healthy nodes and types of malicious nodes are stored based on the type of attacks in the network. The proposed method is based on the combination of feature selection based on Whale optimization algorithm and genetic algorithm with KNN classification in terms of accuracy criteria, has better results than other previous methods. Based on this, it can be said that the Whale optimization algorithm and the genetic algorithm have extracted the features related to the class label well, and the KNN method has been able to well detect the misconduct nodes in the intrusion detection data set in wireless networks.

**KEYWORDS:** Intrusion detection system (IDS), Feature subset selection, Whale optimization, Genetic algorithm, Sample-based classification.

## 1. INTRODUCTION

Wireless networks, are infrastructure-free networks that consist of a set of mobile or fixed hosts that are connected to each other via wireless connections. Each node can act as a final system; in addition it can send and receive bounds in the role of a router [1]. In a wireless network, two nodes may be connected in one step or in several steps. When a source node intends to transfer information to a destination node, bounds are transferred between intermediate nodes, so it is very important for wireless networks to search and quickly create a route from source to destination node [2]. The topology of mobile wireless networks may change intermittently as nodes move, so with these technology nodes can be easily moved with local neighbors. Devices used in wireless networks may exist in different forms but have the same basic activity, i.e., all nodes operate at least to some extent independently [3, 2, 4, 5, 6].

Increasing the need to use equipment such as mobile phones, etc. has become possible with the advent of wireless networks. In the meantime, one of the issues that has always been considered is the issue of attacks on wireless networks [7]. For example, some servers are attacked, causing data to crash and be stolen. In a such situation, substantial and irreparable damage will be done to the network. When it comes to wireless networks, the issue of security and attack prevention becomes much more complex, and this problem is far more serious in the case of mobile wireless networks [8]. Researchers have come up with several ways to solve the problem of security in wireless networks. Approaches are designed to provide security in routing operations and prevent attack threats such as authentication, resource restriction, honesty, confidentiality and protection of users' privacy [9]. Network intrusion and attacks on wireless networks can take many forms: bounds loss, routing structure reconstruction, network topology deviation, and the creation of fake nodes, are types of wireless network attacks [10].

Various methods have been proposed to provide security in wireless networks, but these methods do not have the ability to detect more attacks. Penetration of malicious nodes between nodes in wireless networks may weaken or destroy transmitted data bounds and destroy the entire bounds or part of the available data. Therefore, the overall performance of the network and system is threatened and the results may be interpreted differently [11, 12]. Therefore, preventing and detecting intrusions and attacks in wireless networks has been considered as a fundamental and serious challenge. Thus, network intrusion detection systems (NIDS) [13, 11] are very effective in ensuring security in wireless networks so that these systems are able to cover a wide range of attacks [14, 15].

On the other hand, due to the energy limitation in wireless nodes [16]especially sensors, the use of monitoring equipment such as nodes, for regular monitoring of wireless networks [17, 18]in order to detect intrusion and prevent attack is practically impossible to do [19]. In order to provide solutions to this problem, the issue of remote monitoring systems has recently become one of the most important issues in the field of network security. Remote monitoring of node behavior and how data is sent and received over wireless networks has the potential to detect intrusion and detect attacks on the network. Control systems can also predict future attacks on the network by creating patterns of the behavior [20, 21]. Of course, by detecting malicious nodes, intrusion and attacks in the network can be prevented, which can be considered a threat to the information transmitted in the network. In addition, remote monitoring reduces energy consumption in wireless nodes and is more cost-effective [22, 23].

## 2. Previous methods

Intrusion detection systems are an important tool for network protection. Intrusion detection systems analyze node entry paths in relation to protected systems and specify whether these input paths contain attack nodes. If the intrusion detection system detects an attack, it raises the alert. Traditionally, intrusion detection systems use in-depth bound inspections or situational protocol analysis to detect attacks in network traffic. In 2020, Vijayanand et al. Proposed a Wapper-based method using the modified Whale Optimization (WOA) algorithm. One drawback of WOA is that early convergence leads to an optimal local solution. To overcome this limitation, this paper proposes a method in which genetic algorithm operators are combined with WOA. The fit operator has been used to further improve the whale search space, and the flow operator has helped to avoid getting stuck in the local optimization [6]. The proposed method selects informative features in network data, which help to accurately detect intrusion. Using Support Vector Machine (SVM) [24], they have identified the types of intrusions based on the selected features. The performance of the improved method has been analyzed using the standard CICIDS 2017 and ADFA LD datasets. The proposed method has a better attack detection speed than standard WOA and other evolutionary algorithms. It is also very accurate and suitable for IDS in wireless mesh networks. IDS performance was higher than standard WOA detection by selecting informative features with an improved Wall optimization algorithm. In 2020, Haghnegahdar et al. Developed a new intrusion detection model that can classify cyber-attacks and power system incidents into binary, triple, and multi-class. The intrusion detection model is based on a trained Whale Optimization (WOA) algorithm and artificial neural network (ANN). WOA is applied for initialization and adjustment of the ANN weight vector to achieve the minimum square mean error. The proposed WOA and ANN model can address the challenges of attacks, fault detection and forecasting [25, 26] in the system. The University of Mississippi Laboratory System and Oak Ridge National Laboratory Database Attacks were used to illustrate the proposed model and the experimental results. WOA is able to train

ANN to find the desired weights. The proposed model is used with other common classifiers. Also, a new intrusion detection system based on a combination of a multilayer neural network (MLP) and an artificial bee colony (ABC) and fuzzy clustering algorithms is proposed. Clustering [27]is a process by which a set of objects can be separated into separate groups [28, 29]. In 2018, Wang et al. Introduced a network intrusion detection system with machine learning algorithm to achieve better accuracy and faster detection speed to detect or prevent network attacks. The use of machine learning, and in particular deep learning [30, 31], is another major advantage in which advanced knowledge is not required as much as the blacklist model. Normal and abnormal bounds of network traffic are detected by the MLP, while MLP training is performed using the ABC algorithm with Weight values and bias links are optimized [32]. CloudSim emulator and NSL-KDD dataset are used to test the proposed method. Mean absolute error (MAE), mean square error (RMSE) and kappa statistical parameters are considered as evaluation criteria. The results show that the proposed method is superior to modern methods. Intense Learning Machines (ELMs) are monolayer artificial neural networks that do not need to be replicated. Therefore, their learning speed is fast and speed is very important in the success of network intrusion detection systems and they defend fast and effective response.

### 3. Suggested Method

As mentioned in this research, for the network intrusion detection system based on the combination of selection, a feature set based on the combination of Whale optimization algorithm (WOA) and genetic algorithm with sample-based classification is presented. The proposed method will use the training data set obtained from the KDD Cup data set to determine patterns of network intrusion detection and the test data set prepared from this data set to evaluate the model. In the proposed method, considering the diversity and number of features of user behavior and network traffic, the selection of a subset of features in order to increase the accuracy of model classification, it seems necessary. The purpose of selecting a subset of features is to remove unrelated features and attributes and plugins so that in addition to reducing the data dimension and reducing the operational and spatial complexity of the system, the classification accuracy can be increased and high speed and lower cost data classification. In addition, the selection of a subset of features can detect the implicit dependence between the data and the class label of this data so that the classification of test samples that will be added to the model in the future can be easily done. For this purpose, in the proposed method, the WOA GA-based feature subset selection approach has been used to determine the features that are important and related to the class label. In general, having irrelevant and redundant features due to solving the classification problem, affects the classification performance, especially with high-dimensional data sets, and reduces the classification accuracy. In addition, having a large number of features leads to over-training and low ability to generalize the new data and detect test samples. Thus, the task of detecting intrusion in wireless networks, given the high number of nodes and the various features and behavior of each of these nodes in the network, can become a difficult classification task that it cannot bring optimal results .

The purpose of selecting features is to eliminate the difficulty of classification and increase the accuracy of classification by selecting the relevant features. In one-objective feature selection tasks, feature selection has a goal for optimization. The sole purpose of feature selection is to find the best feature combination for optimal classification performance, regardless of training cost or number of features. Feature Selection takes the task of selecting a subset of features by turning them into an optimization problem, in which the goal is to create a balance in order to optimize multiple goals. The goals that this optimization method pursues in order to select a subset of features include reducing the entropy of features based on the class label, increasing the standard deviation for differentiating samples of two classes, and increasing the classification accuracy of nodes in the network, and will increase the accuracy of predicting test samples. In fact, the goals include two sets of reducing the number of features and increasing the classification performance. Consequently, the solution to the property choice optimization problem is a set of dominant solutions, in which each solution is a vector of two components, the number of features and the degree of classification error. By using the feature selection problem as a minimization problem, the goal is to minimize the number of unrelated features and to minimize the classification error rate. In the following, we will formulate the proposed method.

### 3.1. Formulation Problem

In this research, the feature selection problem is considered as a multi-objective optimization problem which is solved using a multi-objective particle swarm optimization algorithm. The goals of this optimization problem are classified into two general groups, one is to reduce the number of features and the other is to reduce the amount of classification error. Accordingly, the number of data set properties, independently, indicates the dimensions of the problem with the binary search space in the range between [0, 1]. Because MOPSO is an initial population-based meta-heuristic approach, the initial population corresponds to a potential subset of features that are directly related to the class tag. As shown in Figure 1, in the decision search space, $x_1$ is a real positive number that indicates the error rate and $x_2$ is a real positive number that indicates the number of attributes, and the function F which leads to a set Balance is made from decision vectors that reduce both the error rate and the number of attributes denoted by [$F(x_1)$, $F(x_2)$]. For example, one decision vector may have values $v_1$ = [0.001, 12] and another vector may have values $v_1$ = [0.05, 5]. If the evaluation function prefers the lowest error rate regardless of a number of features, $v_1$ is naturally a satisfactory solution. If so, $v_2$ is a satisfactory solution for a target function about having a minimum number of attributes. Hence, both $v_1$ and $v_2$ are acceptable solutions, but in the end, another victim becomes a target and returns to the decision-maker's priority.

In the proposed method, the initial population is adjusted with the features in the KDD-CUP dataset. Thus, each particle is a binary vector whose value is equal to the number of properties, and each element points to a property in the data set. In this case, each particle in the swarm indicates the property selection with a value of 1. Hence, the length of a particle is equal to the number of features in the corresponding dataset. Figure 1 shows an example of a particle representation, which shows the number of properties in a data set equal to 41 properties.

| 0 | 1 | 0 | 0 | 0 | 1 | 0 | 1 | ... | 1 |
|---|---|---|---|---|---|---|---|-----|---|
| $F_1$ | $F_2$ | $F_3$ | $F_4$ | $F_5$ | $F_6$ | $F_7$ | $F_8$ | ... | $F_{41}$ |

**Fig.1.** shows the initial population vector of the data set sample

As shown in Figure 1, each particle in the particle swarm optimization algorithm is a set of features in the data set that are randomly the number of components of this vector may be zero and the number 1. Values that have a value of zero represent a property that has not been selected, and in contrast, resources that have a value of one indicate the selection of a property related to that knowledge. As a result, we can see from Figure 1 that the selected subset of features includes the feature set [$F_2$, $F_6$, $F_8$, $F_{41}$].

In the proposed method, in order to select the properties for the initial particle vectors, the Sigmoid transfer function ($S_1$) is used to define the probability of selecting the property or why not selecting it. the function of this function is such that if the random probability is less than the threshold value of the transfer function, which is generally considered to be 0.5, it assigns a value of zero to the value associated with this property. Otherwise, a value of 1 is recorded for this attribute and the attribute will be evaluated based on the objective functions. After selecting the initial population, based on the nature of the particle swarm optimization algorithm, the initial location and velocity of each particle is determined using evaluation functions. The location of each particle in this method is considered as the selected properties of the properties in the data set and the speed of each particle is considered as the speed of convergence to the high classification rate and reduction of classification error. The properties that have the highest value of the evaluation function and the greater particle swarm around them, results as the output of the initial stage of property selection. The best particle position and velocity results are saved at this stage and the particle position is updated. This process continues until we reach a final answer that strikes a balance between the goals.

### 3.2. The objective function

As mentioned, the proposed method uses a multi-objective particle swarm algorithm to select a subset of class tag-related features. In this method, multiple goals are combined and finally we achieve two general categories of features in the form of minimization. The evaluation of selected subsets of features is based on the two main objectives of reducing the number of features and the classification error rate. In order to evaluate the initial population and select the expert population and find the particles with the highest weight, the proportionality function is expressed as Equation.1.

$$Minimize\ F(x) = \begin{cases} f_1(x) = \frac{L}{A} & , L \in A, A \in \mathbb{R}^+ \\ f_2(x) = 1 - \frac{FP+FN}{P+N} & , (P+N) \in \mathbb{R}^+ \end{cases} \quad Eq.1$$

In relation L, the number of features selected from the data set and A is the total number of features. In order to evaluate the error rate of each particle according to the selected properties in each step, the perturbation matrix criteria in which the true positive (TP) is used to represent those normal nodes that are correctly represented by the classification model as normal according to the selected features. False positive (FP) to indicate those normal nodes that have been misdiagnosed by the classification model and intrusion according to the selected characteristics, true negative (TN) to indicate those Infiltration nodes that are correctly classified by the model and according to the selected characteristics, the intrusion is detected and finally the real positive (FN) to show those intrusion nodes that are mistaken by The classification model is applied and normalized according to the selected features. In Equation 1, P is equal to the sum of TP+FN and N is equal to FP+TN. The first objective function $f_1$ (x) is related to the ratio of the selected attributes to the total attributes in the data set, while the second objective is $f_2$ (x) to evaluate the classification error rate.

In the proposed method, the particles at each stage use the SVM algorithm to evaluate the training samples in order to evaluate the amount of classification error for each particle and by obtaining the classification error rate, according to the number of features selected, the best particles in each step are selected and ranked in each step based on the degree of optimality. By converting these properties obtained from each particle into property vectors, SVM will try to differentiate between healthy nodes and malicious nodes and create a safe margin between the two classes. Finally, the least amount of error and the least number of features selected from the training data set will be selected as the best particle and the extracted features will be specified as the classification pattern. Figure 2 shows the proposed flowchart.

### 4. Implement the proposed method

In this part of the research, we first describe the standard data set of intrusion in wireless networks. Then, in order to implement the proposed method, we will first implement the feature subset selection method based on the WOA-GA algorithm and then evaluate it using the nearest neighbor k algorithm.

### 4.1. Standard data penetration in wireless networks

Due to the nature of the proposed method, in order to implement the combined feature selection method based on Whale optimization algorithm and Genetic algorithm, we will use a standard network penetration data set called kddcup99 which is stored in the UCI data repository. The present dataset includes the activities of nodes in the wireless network that have been simulated in the Lincoln Lab for several months. By monitoring the nodes in this experiment in order to collect the standard data set, the status of each node in the data set is determined. Depending on the behavior of the nodes in the network, each node is either considered as a healthy node or represents some kind of intrusion into the wireless network. In this regard, in the method of proposing data sets in two classes of healthy nodes and Abuse nodes are trained as network penetration.

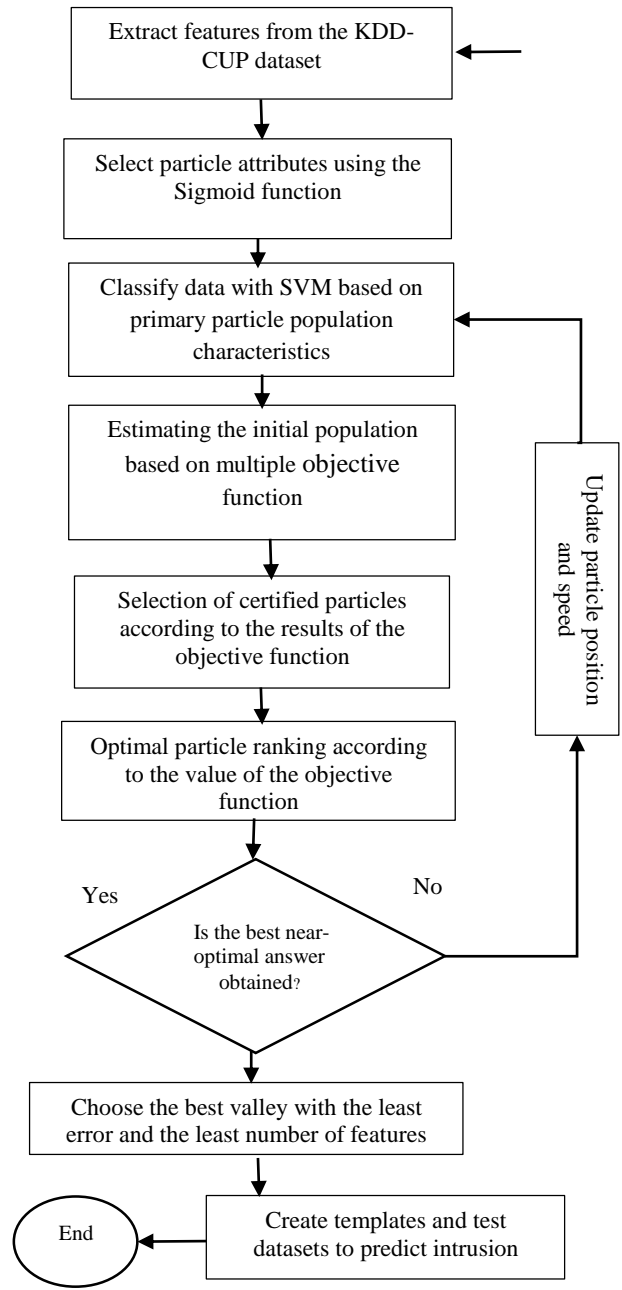

**Fig.2.** the proposed flowchart

### 4.2. Implement WOA GA-based feature selection

According to the proposed method, the initial population in the WOA-GA algorithm is binary vectors, which represent each of the index properties of the desired property in the main data set. The length of the vectors as the initial population is proportional to the number of attributes in the main data set. So each Whale is defined as a solution in a vector with n elements, where n is proportional to the number of attributes in the original data set. In the proposed method, first 100 Whales (solution) are selected as the initial random population. In the WOA-GA algorithm, the initial population consists of a matrix with dimensions of 41 x 100 of random numbers between zero and one. Each element of this matrix in row i and column j represents the probability of the property j in the solution (Whale) of i. According to the random selection function in the proposed method, the values within the components of this matrix are binary. In this way, if the value of each drive is less than the threshold, the drive will be set to zero, otherwise it will be set to 1 value. In the initial population matrix, the j property will not exist in the i solution if the $A_{ij}$ array has a value of zero, but if the value of this array is equal to one, the j property is one of the properties selected in the solution properties of i subset in (Whale).

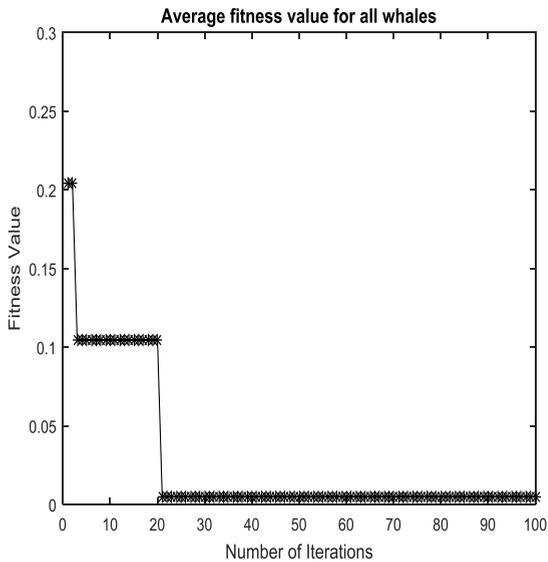

**Fig.3.** Mean Proportional Values for Solutions in WOA

As can be seen from Figure 3, the proportionality values for the solutions decreased during the iteration, from 0.024 to 0.005 per 100 iterations of the WOA-GA algorithm. The mean values of the proposed solution indicate the superiority of the WOA-GA algorithm in finding a subset of features with the least amount of error.

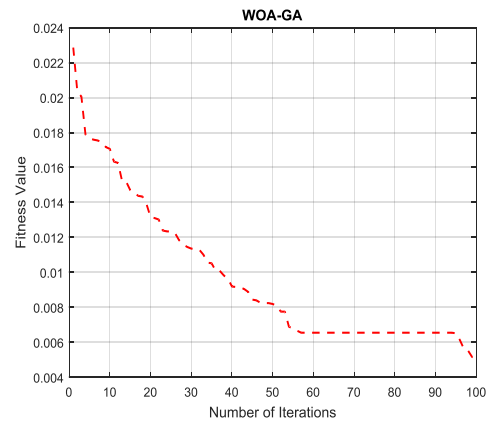

**Fig.4.** Convergence of the WOA-GA algorithm towards the optimal point

According to Figure 4, what can be seen is that the WOA-GA algorithm has a rapid convergence to the optimal point, which is zero. Given that the nature of the fit function in the proposed method of minimizing the value of the fit function and minimizing the amount of intrusion detection error in the wireless network, therefore the rapid convergence to the minimum point indicates the correct choice instead of combining the optimization pattern Whale and Genetic algorithms in order to select a subset of useful features in the standard network intrusion detection dataset.

**Table1.** Subset of features selected by the WOA-GA algorithm

| Feature No. | Feature Name | Feature No. | Feature Name |
|---|---|---|---|
| 1 | "Duration" | 17 | "Num_file_Creations" |
| 2 | "Protocol_Type" | 18 | "Num_shells" |
| 5 | "Dst_bytes" | 19 | "Num_access_files" |
| 6 | "Flag" | 23 | "Serror_rate" |
| 10 | "Hot" | 29 | "Srv_rerror_rate" |
| 12 | "Logged_in" | 32 | "dst_host_srv_count" |
| 13 | "Num_compromised" | 34 | "dst_host_diff_srv_rate" |
| 14 | "Root_shell" | 39 | "dst_host_rerror_rate" |
| 15 | "Su_attempted" | 40 | "dst_host_srv_rerror_rate" |
| 16 | "Num_root" | 41 | "Is_hot_login" |

## 5. Evaluate the proposed method

Evaluation of the proposed method to analyze and evaluate the performance of the proposed method on the test data set, which is part of the initial standard data set, is in order to identify new attacks in the network. Therefore, different criteria have been introduced to evaluate the performance of network intrusion detection systems, each of which examines a different aspect of the model, but in general, the improvement of each criterion indicates an improvement in the model performance in the standard intrusion detection dataset in wireless networks. In this study, the criteria related to the turbulence matrix have been used to evaluate the proposed method. Thus, based on the confusion matrix, the results of this comparison can be summarized in four parameters:

- True positive TP: nodes that are properly selected as healthy nodes based on the proposed model.
- The real negative of FP: is the nodes that have been correctly selected as abusive nodes based on the proposed model.
- False Positive TN: nodes that have been selected as healthy nodes based on the proposed model, but are in fact a malicious node.
- True positive FN: nodes that have been selected as abusive nodes based on the proposed model, but are in fact a healthy node.

Given that the confusion matrix is a well-known mechanism for evaluating the classification performance in two-part datasets (healthy nodes versus abusive nodes), evaluation criteria can be derived from the confusion matrix. These criteria include accuracy, sensitivity, precision and F-criteria.

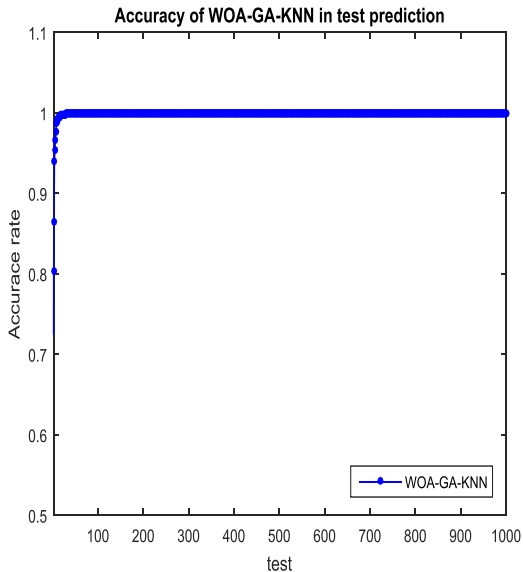

**Fig.5.** Accuracy diagram of WOA-GA algorithm for test data set

According to Figure 5, what can be seen is that the WOA-GA algorithm for selecting the set of effective features and combining it with the KNN algorithm has high accuracy standard values for detecting malicious nodes in the test data set and serious nodes which join the pseudo. Also, according to Figure 6, what can be seen is that the WOA-GA algorithm for selecting the effective feature subset and combining it with the KNN algorithm has high sensitivity criteria for detecting malicious nodes in the test data set and serious nodes which join the pseudo.

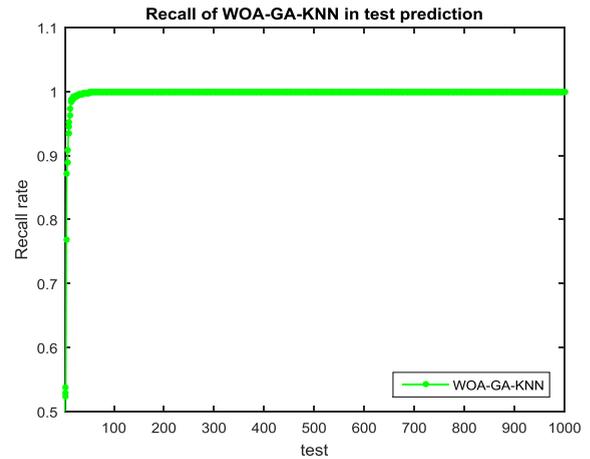

**Fig.6.** Sensitivity diagram of WOA-GA algorithm for test data set

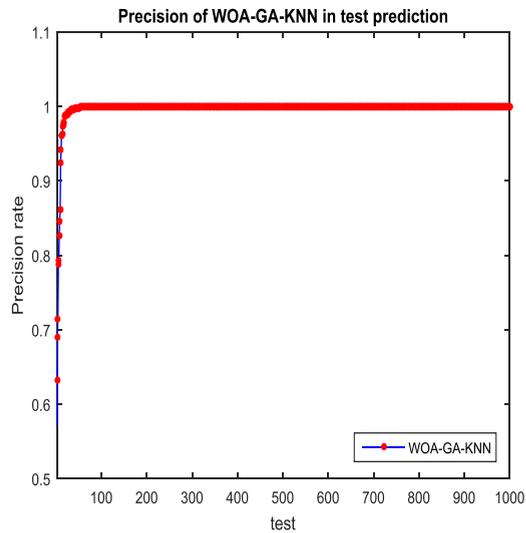

**Fig.7.** Diagram of the accuracy of the WOA-GA algorithm for the test dataset

According to Figure 7, what can be seen is that the WOA-GA algorithm for selecting the effective feature subset and combining it with the KNN algorithm has high accuracy standard values for detecting abusive nodes in the test data set and serious nodesw which join the pseudo.
According to Figure 8, what can be seen is that the WOA-GA algorithm for selecting the effective feature subset and combining it with the KNN algorithm has high F-value values for detecting malicious nodes in the test data set and serious nodes which join the pseudo.

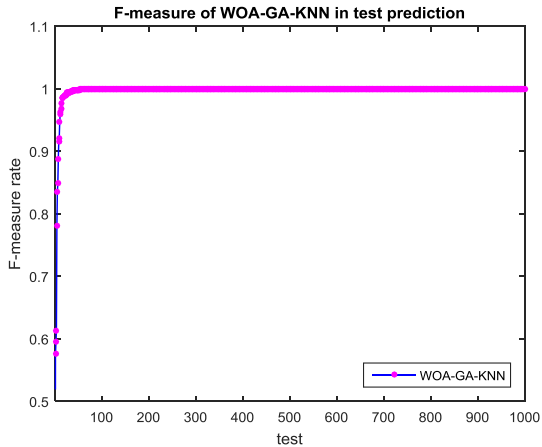

**Fig.8.** Benchmark of WOA-GA algorithm for test data set

## 6. Comparison of the proposed method with previous methods

After evaluating the performance of the proposed method against the test data set, in order to check the accuracy of the results of the proposed method, which is a combination of Whale optimization algorithm and genetic algorithm and KNN classification, the proposed method in terms of node detection accuracy Abuse in the network of previous methods [7, 33] will be compared in this regard. Figure 9 Comparison of the proposed method with previous methods is illustrated in terms of accuracy criteria.

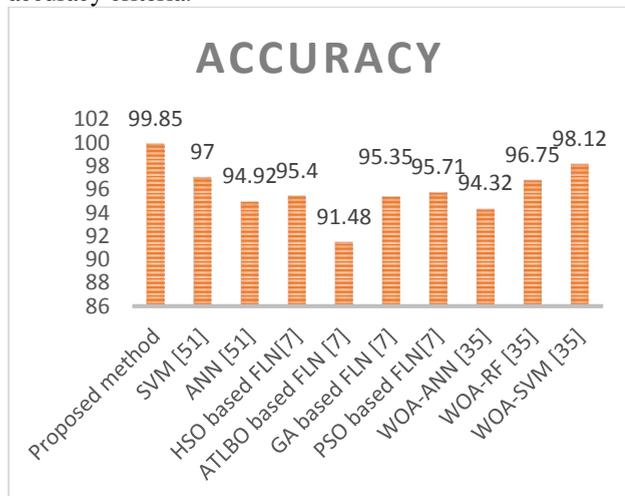

**Fig.9.** Comparison of the proposed method with previous methods

According to Figure 9, what can be seen is that the proposed method has been compared with the previous methods according to the accuracy criterion. According to Figure 9, it can be seen that the proposed method based on the combination of feature selection based on Whale optimization algorithm and Genetic algorithm with KNN classification in terms of accuracy criteria, has better results than other previous methods. Based on this, it can be said that the Whale optimization algorithm and the Genetic algorithm have extracted the features related to the class label well, and the KNN method has been able to well detect the misconduct nodes in the intrusion detection data set in wirless-networks.

## 7. Conclusion

Due to the widespread use of communication networks and the ease of communication through wireless networks, this type of network has received more attention than ever before. Ability to use in any environment without the need for environmental monitoring and engineering of these networks, has increased their use in various fields. Increased use of them has also raised security issues in the field of sending and receiving information, which is known as intrusion detection as the most important issue. Network intrusion detection is the process of identifying malicious activity in a network by analyzing network traffic behavior. These networks are relatively vulnerable to other networks due to their wireless nature, limited resources, mobility and dynamics, and their important and vital tasks. There are several ways to provide security in wireless networks, but these methods are not able to detect more attacks. In addition, due to the limited energy of wireless nodes, it has become virtually impossible to use monitoring nodes for permanent monitoring in wireless networks in order to prevent and detect intrusions and attacks in these types of networks. Therefore, intrusion detection and network detection systems (NIDS) play an important role in providing security in wireless networks and can cover a wide range of attacks. In this research, a network intrusion detection system using feature selection based on a combination of Whale optimization (WOA) and Genetics (GA) and sample-based classification is proposed. In this research, the standard data set KDDCUP1999 has been used in which the characteristics related to healthy nodes and types of malicious nodes are stored based on the type of attacks in the network. The proposed method is based on the combination of feature selection based on Wall optimization algorithm and genetic algorithm with KNN classification in terms of accuracy criteria, has better results than other previous methods. Based on this, it can be said that the Whale optimization algorithm and the genetic algorithm have extracted the features related to the class label well, and the KNN method has been able to detect the nodes of misconduct in the data set of intrusion detection in wireless networks.